\documentclass[preprint,floatfix,aps,showpacs]{revtex4}
\newcommand {\bea}{\begin{eqnarray}}
\newcommand {\eea}{\end{eqnarray}}
\newcommand {\be}{\begin{equation}}
\newcommand {\ee}{\end{equation}}

\usepackage{graphicx}
\usepackage{epsfig,subfigure}
\usepackage{amsmath}
\begin{document}
\def\({\left(}
\def\){\right)}
\def\[{\left[}
\def\]{\right]}

\def\Journal#1#2#3#4{{#1} {\bf #2}, #3 (#4)}
\def\RPP{{Rep. Prog. Phys}}
\def\PRC{{Phys. Rev. C}}
\def\PRD{{Phys. Rev. D}}
\def\FP{{Foundations of Physics}}
\def\ZPA{{Z. Phys. A}}
\def\NPA{{Nucl. Phys. A}}
\def\JPG{{J. Phys. G }}
\def\PRL{{Phys. Rev. Lett}}
\def\PRpt{{Phys. Rep.}}
\def\PLB{{Phys. Lett. B}}
\def\AP{{Ann. Phys (N.Y.)}}
\def\EPJA{{Eur. Phys. J. A}}
\def\NP{{Nucl. Phys}}  
\def\ZP{{Z. Phys}}
\def\RMP{{Rev. Mod. Phys}}
\def\IJMPE{{Int. J. Mod. Phys. E}}

\input epsf

\title{Criteria for nonlinear parameters of relativistic mean field models}

\author{A. Sulaksono$^{1,4}$, T. J. B{\"u}rvenich$^{2}$, P. -G. Reinhard$^{3}$,  and J. A. Maruhn$^{4}$}

\affiliation{$^1$Departemen Fisika, FMIPA, Universitas Indonesia,
Depok, 16424, Indonesia\\
$^2$Frankfurt Institute for Advanced Studies, Universit\"at Frankfurt, 60438 Frankfurt am Main, Germany\\
$^3$Institut
f\"ur Theoretische Physik II, Universit\"at Erlangen-N\"urnberg, D-91058 Erlangen, Germany\\
$^4$Institut
f\"ur Theoretische Physik,Universit\"at Frankfurt, 60438 Frankfurt am Main, Germany }

\begin{abstract}
Based on the properties of the critical and the actual effective
masses of sigma and omega mesons, criteria to estimate the values of the
isoscalar nonlinear terms of the standard relativistic mean field model
that reproduce stable equations of state in respect to particle hole
excitation at high densities are derived. The relation between nuclear
matter stability and the symmetric nuclear matter properties are
shown. The criteria are used to analyze in a more systematic way  the
high-density longitudinal and transverse instabilities of some
parameter sets of relativistic mean field models. The critical role of
the vector and vector-scalar nonlinear terms is also discussed quantitatively.

\end{abstract}
\maketitle
\newpage

%%%%%%%%%%%%%%%%%%%%%%%%%%%%%%%%%%%%%%%%%%%%%%%%%%%%%%%%%%%%%%%%%%%%%%%%%
\section{Introduction}
\label{sec_intro}
Relativistic mean field (RMF) models have been quite successful in
providing a microscopic description of many ground-state properties
from medium to heavy nuclei. The parameters of each model are obtained
by adjusting its parameters to a few ground-state properties of a set
of magic and semi-magic nuclei. The differences between one parameter
set and the others are not only due to differences in choosing
observables, strategies, and constraints for the parametrization, but
also to the difference in the nonlinear ansatz used. Even in recent
years, there have still been some efforts to generate new RMF
parameter sets~\cite{Lala09,Sharma08,Kumar,Pieka2}. The aim is to
improve and to extend the applicability of the model for the
description not only of ground states but also of excited states of
finite nuclei and nuclear matter. Indeed, even the boundaries are not
too clear, the range of applicability of RMF models is
limited. Therefore, applying the models in regions which are outside
their fitting window by extrapolation needs special care.

Recently, considerable attention has been paid in constraining the
equation of state (EOS) of matter beyond the nuclear matter saturation
density ($\rho_0$) using flow data from heavy-ion collision
experiments and astrophysical observatories~\cite{Klahn, Sagert,
Pieka07}. In order to be consistent with data, the EOS should be
relatively soft at moderate densities ( $\le5 \rho_0$) and stiff at
high densities ( $>5 \rho_0$). Many models, including some
parameter sets of RMF models have been checked. The author of
Ref.~\cite{Pieka07}, also used some model independent calculations of
pure neutron matter as an additional theoretical constraint to test
the validaty of two parameter sets of RMF models (NL3 and FSUGold) at
high densities.  It is found that FSUGold is consistent with all
constraints except for a high density EOS that appears mildly softer
than required by astronomical observations~\cite{Pieka07}. It seems
that some parameter sets of RMF models are unable to fulfil the above
requirements~\cite{Klahn, Sagert, Pieka07}. At this point, we have
sufficient reason to re-investigate the parameterizations of RMF
models with high density applications in mind.

We studied the high density instabilities of some representative
parameter sets of RMF models with respect to density fluctuation by
observing their longitudinal and transverse particle-hole excitation
modes~\cite{AMBM07}. We have found that certain parameter sets are
unstable at high densities but for some parameter sets with additional
vector and vector-scalar nonlinear terms, the onset of instabilities
can be pushed into a region with quite large matter density ($\rho_B$)
and perturbed momenta ($q$).  It means that the longitudinal and
transverse parts of the particle-hole excitation modes of
relativistic mean field models depend sensitively on the isoscalar
nonlinear terms used. Therefore by adjusting these nonlinear terms,
the instabilities at high densities of RMF models can be avoided.

In this work, we extend the results of our previous
letter~\cite{AMBM07} by deriving criteria to estimate the values of
isoscalar nonlinear terms of the standard RMF model (RMF model with
minimal nonlinear terms) that produce a stable EOS at high
densities. We then employ these criteria to systematically study the
high density longitudinal and transverse instabilities of some
representative RMF models. In this way, the actual role of vector and
scalar-vector nonlinear terms can be revealed in a more quantitative
manner. These analyses can provide a practical hint that should be
useful for RMF parameterization.  The extension to the isovector
sector is also important. The question whether the parameter set of
the standard RMF model with stable nuclear matter EOS at high
densities is consistent with the above constraints and simultaneously
has acceptable predictions for finite nuclei, would also be very
interesting to investigate, but these points need additional
considerations that are beyond the scope of this work and will left to
a future paper.

This paper is organized as follow: in Sec.~\ref{sec_constraint} it is
discussed how to estimate the isoscalar nonlinear parameter sets of
the standard RMF model; in Sec.~\ref{sec_stability} we discuss the
stability of some RMF models; and finally in Sec.~\ref{sec_conclu} the
conclusions are drawn.
  
%%%%%%%%%%%%%%%%%%%%%%%%%%%%%%%%%%%%%%%%%%%%%%%%%%%%%%%%%%%%%%%%%%%%%%%%%
\section{Constraining nonlinear parameters of standard relativistic mean field models}
\label{sec_constraint}

The instability of nuclear matter at high density is generated by
particle-hole excitation modes. These modes are indicated by the
existence of poles of the meson propagators at zero energy
transfer.~\cite{Horo3,Fri,PGR2}.  In this section, the criteria for
the isoscalar nonlinear parameter values of the relativistic mean
field model that cause instabilities at relatively high densities are
investigated.

We start from the energy density $\varepsilon$ of standard RMF models
in symmetric nuclear matter (SNM) as,
\bea
\varepsilon&=&\varepsilon_{\rm linear}+\frac{1}{3} b_2\sigma^3+ \frac{1}{4} b_3\sigma^4-\frac{1}{4} c_3  V_0^4 ,
\label{eq:sd_edens}
\eea
where $\varepsilon_{\rm linear}$ is the energy density of the linear
Walecka model.  Note that from now on, we denote parameter sets with
nonlinear vector parameter $c_3$=0 as S-RMF, and those with this
parameter not equal to zero as V-RMF. The transverse
($\epsilon_T$) and longitudinal ($\epsilon_L$) dielectric functions at
$q_0=0$ for SNM are~\cite{AMBM07}
\bea
\epsilon_T&=&1+2 d_V^T \Pi_T,\nonumber\\\epsilon_L&=&1+2 d_S
\Pi_S-2 d_V^L \Pi_V,
\label{eq:die}
\eea
with the effective vector polarization defined as
\bea
\Pi_V &\equiv& \Pi_{00}-2 d_S\Pi_M^2+ 2d_S\Pi_S  \Pi_{00}.
\eea
The explicit forms of the polarizations are 
\begin{widetext}
\bea
 \Pi_{T}(q,0)&=& \frac{1}{ \pi^2} \bigg{\{}\frac{1}{6} k_F E_F - \frac{q^2}{6}  ~{\rm ln} \frac{k_F+E_F}{M^*}\nonumber\\
&-&  \frac{E_F}{6 q}(2 M^{*2}- k_F^2-\frac{3}{4}q^2) {\rm ln} ~\Big| \frac{2 k_F - q}{2 k_F + q} \Big| -\frac{ E}{3 q}(M^{*2}-\frac{1}{2}q^2)  {\rm ln} ~\Big| \frac{q E_F - 2 k_F E}{q E_F + 2 k_F E} \Big| \bigg{\}},\nonumber\\
\eea
\end{widetext}
for the transverse polarization,
\begin{widetext}
\bea
 \Pi_{S}(q,0)&=& \frac{1}{2 \pi^2} \bigg{\{} k_F E_F - (3 M^{*~2} + \frac{q^2}{2})  ~{\rm ln} \frac{k_F+E_F}{M^*}\nonumber\\
&+&  \frac{2 E_F E^2}{q} {\rm ln} ~\Big| \frac{2 k_F - q}{2 k_F + q} \Big| -\frac{2 E^3}{q}  {\rm ln} ~\Big| \frac{q E_F - 2 k_F E}{q E_F + 2 k_F E} \Big| \bigg{\}},
\eea
\end{widetext}
for the scalar polarization, 
\bea
\Pi_{M}(q,0)= \frac{M^*}{2 \pi^2}  \bigg{\{} k_F - (\frac{ k_F^2}{q}-\frac{q}{4}) {\rm ln} ~\Big| \frac{2 k_F - q}{2 k_F + q} \Big|\bigg{\}},
\eea
for the mixed scalar-vector polarization, and 
\begin{widetext}
\bea
\Pi_{00}(q,0)&=&-\frac{1}{ \pi^2} \bigg{\{}\frac{2}{3}  k_F E_F - \frac{q^2}{6} ~{\rm ln} \frac{k_F+E_F}{M^*}
 - \frac{E_F }{3 q} ( M^{*~2} + k_F^2-\frac{3 q^2}{4})  {\rm ln} ~\Big| \frac{2 k_F - q}{2 k_F + q} \Big|\nonumber\\ &+&\frac{E}{3 q}(M^{*~2} - \frac{q^2}{2}) {\rm ln} ~\Big| \frac{q E_F - 2 k_F E}{q E_F + 2 k_F E} \Big| \bigg{\}},
\eea
\end{widetext}
for the longitudinal polarization, with Fermi momentum $k_F$, nucleon
effective mass $M^* = M - g_{\sigma} \sigma$, Fermi energy $E_F$=
$(k_F^2+M^{* ~2})^{1/2}$, and $E $=$(q^2/4+M^{* ~2})^{1/2}$.

The longitudinal scalar meson propagator is given by
\bea
d_S=\frac{g_{\sigma}^2}
{q^2+m^{*~2}_{\sigma}},
\label{eq:s_prop}
\eea 
while the vector meson longitudinal and transverse propagators are
\bea
d_V^L=d_V^T =\frac{g_{\omega}^2}{q^2+m^{*~2}_{\omega}},
\label{eq:v_prop}
\eea 
where the $\sigma$ and $\omega$ mesons effective masses are given by
\bea
m_\sigma^{*~2}&=& \frac{\partial^2 \epsilon}{\partial^2 \sigma}=m_\sigma^2+2 b_2 \sigma+3 b_3 \sigma^2,\nonumber\\ m_\omega^{*~2}&=& -\frac{\partial^2 \epsilon}{\partial^2 V_0}=m_\omega^2 + 3 c_3  V_0^2 .
\label{eq:meseffmass}
\eea

If we define $m_\sigma^{*~2~c}$ and $m_\omega^{*~2~c}$ as
$m_\sigma^{*~2}$ and $m_\omega^{*~2}$ at $\epsilon_T$=0 and
$\epsilon_L$=0 simultaneously, then from Eqs.~(\ref{eq:die},
\ref{eq:s_prop} and \ref{eq:v_prop}) we have
\bea
m_\omega^{*~2~c}&=&-[2 g_{\omega}^2 \Pi_T + q^2],\nonumber\\
m_\sigma^{*~2~c}&=&-\[\frac{2 g_{\sigma}^2 (\Pi_S \Pi_T-\Pi_M^2+\Pi_S\Pi_{00})}{( \Pi_T+\Pi_{00})}+q^2\].
\label{eq:msqmass}
\eea
Because the presence of poles in the meson propagators is equivalent
with $\epsilon_T$=0 and $\epsilon_L$=0, it implies that at any nuclear
matter density and momentum $q$, the stable regions are determined
from the following criteria
\bea
 \frac{m_\sigma^{*~2~c}}{m_\sigma^{*~2}}< 1,\nonumber\\
 \frac{m_\omega^{*~2~c}}{m_\omega^{*~2}}< 1.
\label{eq:requr}
\eea 

We observe that the $m_\sigma^{*~2~c}$ and $m_\omega^{*~2~c}$ of every
RMF parameter set strongly depend on the values of $g_{\sigma}$ and
$g_{\omega}$ but are insensitive to the values of the parameters
$b_2$, $b_3$, and $c_3$.  In the contrary, $m_\sigma^{*~2}$ and
$m_\omega^{*~2}$ do not depend explicitly on $g_{\sigma}$ and
$g_{\omega}$ values but they are very sensitive to the values of
$b_2$, $b_3$, and $c_3$. Fortunately, we also observe that most of the
standard RMF parameter sets in the literature's with acceptable
predictions for nuclear matter and finite nuclei have similar
$g_{\sigma}$ and $g_{\omega}$ values. Let us take the average values
of both coupling constants as the ``standard'' values of both
parameters.  We found that, for any parameter set, if the values of
these $g_{\sigma}$ and $g_{\omega}$ parameters are larger than the
standard ones, its $m_\sigma^{*~2~c}$ and $m_\omega^{*~2~c}$ are
higher and on the other hand, if these coupling constants are smaller,
its $m_\sigma^{*~2~c}$ and $m_\omega^{*~2~c}$ are lower.

\begin{figure*}
\epsfig{figure=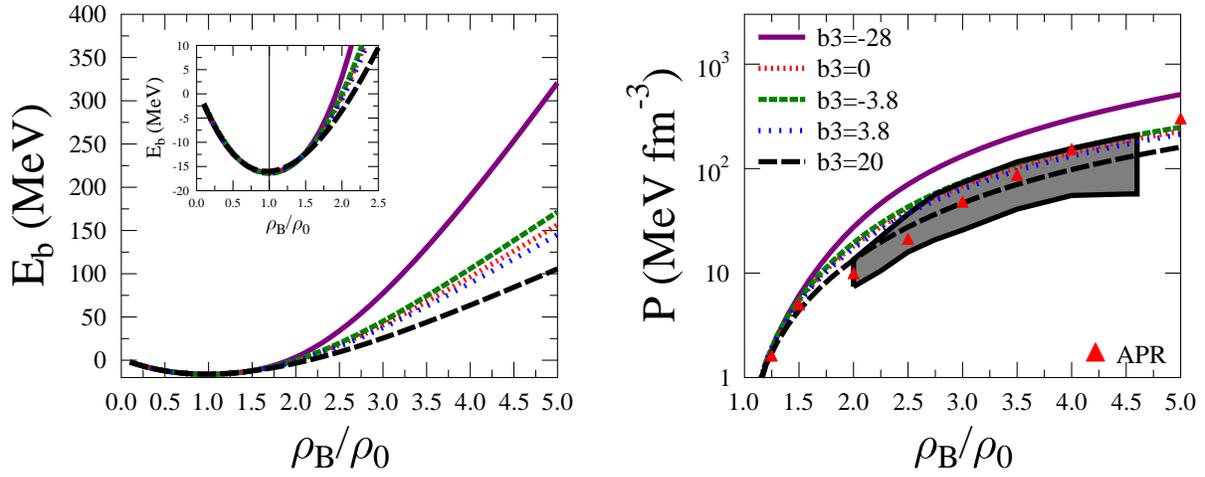, width=18cm}
\caption{The SNM binding energy (left panel) and pressure (right
panel) calculated using test parameter sets in
Table~\ref{tab:param_trmf}. The Shaded region in the right panel
corresponds to experimental data of Ref.~\cite{Daniel}, and the triangles to data calculated from the microscopic model of
Ref.~\cite{Apr}. The inset figure shows the low-density region at
better resolution.}
\label{t_eos}
\end{figure*}

\begin{figure*}
\epsfig{figure=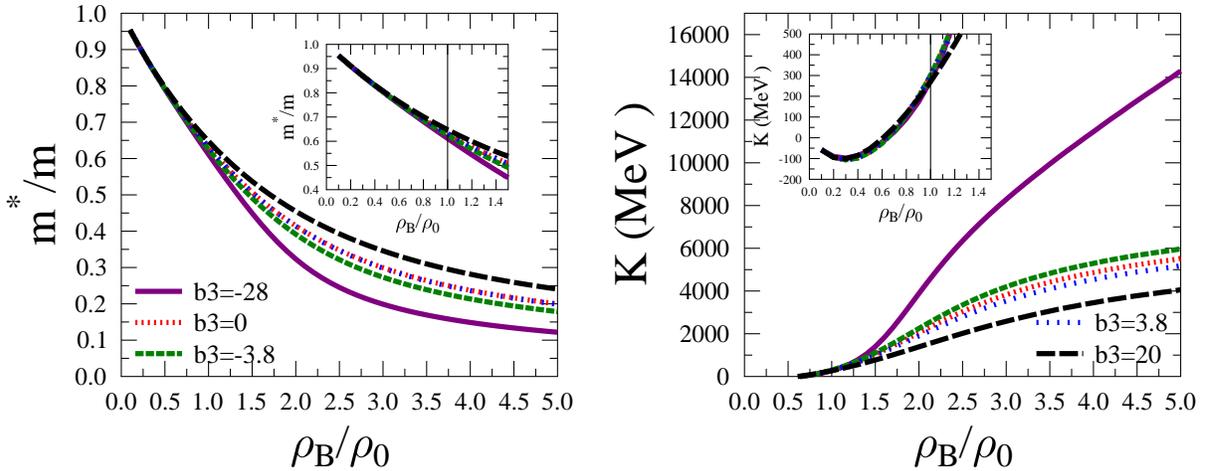, width=18cm}
\caption{The SNM nucleon effective mass (left panel) and
compressibility (right panel) calculated using the test parameter sets
of Table~\ref{tab:param_trmf}. The inset figure shows the low-density
region at better resolution.}
\label{t_effmass}
\end{figure*}

Based on the fact that almost all of the standard RMF parameter sets
in the literature have similar values of $g_{\sigma}$ and
$g_{\omega}$, we will use Eq.~(\ref{eq:requr}) to determine a lower
limit for the $b_2$, $b_3$, and $c_3$ parameters of standard RMF
models that are stable with respect to the particle-hole excitation
modes at high densities. For the transverse mode, it is relatively
straightforward to extract one parameter $c_3$ from a critical
effective omega meson mass, but for the longitudinal one, to pick up
two parameters $b_2$ and $b_3$ from a critical effective sigma meson
mass can generate many combinations because the correlation between
$b_2$ and $b_3$ for each parameter set is different. It depends on
many factors, like the values of other parameters, fitting procedure
and strategy including the choice of observables for parameterization
etc. To avoid this problem, we generate some test parameter sets with
almost the same value of $g_{\sigma}$ and $g_{\omega}$ with variation
of $b_3$ ( the $b_3$ term dominates over the $b_2$ term at high
densities) while the $b_2$ and $c_3$ parameters are adjusted such that
the standard SNM properties at saturation are fulfilled. The
parameters of some exemplary test parameter sets can be seen in
Table~\ref{tab:param_trmf}. The SNM binding energy and pressure
predicted by each test parameter in this table are displayed in
Fig.~\ref{t_eos}, while their ratioes of effective mass to nucleon
mass and the compressibilities are shown in Fig.~\ref{t_effmass}. It
is clear from both pictures that they have similar nuclear matter
properties at saturation (insets) but are different at high
densities. The EOS of parameter sets with $b_3\ge 0$ tends to be
closer to the one of the experimental data from Ref.~\cite{Daniel} and
the microscopic calculation of Ref.~\cite{Apr}.

\begin{figure*}
\epsfig{figure=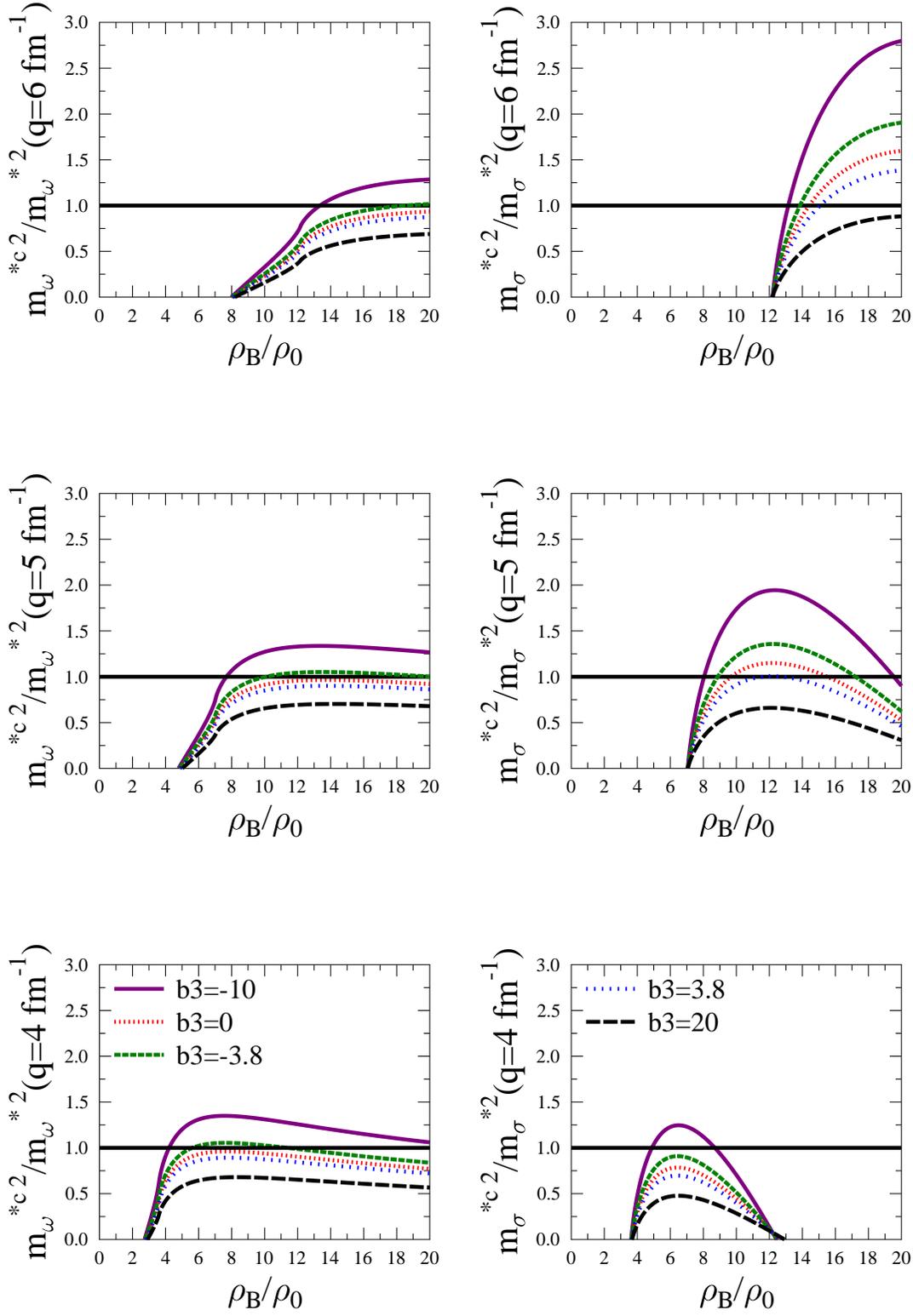, width=16cm}
\caption{ $m_\sigma^{*~2~c}/m_\sigma^{*2}$ and
$m_\omega^{*~2~c}/m_\omega^{*2}$ as functions of $\rho_B/\rho_0$ 
for various values of $q$ and $b_3$ as indicated.}
\label{crit_mass_trmf}
\end{figure*}
\begin{table}
\centering
\caption {Isoscalar parameters for some test parameter sets. }
\label{tab:param_trmf}
\begin{tabular}{c c c c c c c}
\hline\hline ~Parameter~~&~test1  & test2   &test3 & test4 &test5  \\\hline
$g_{\sigma}$ & 10.08 & 10.18 & 10.02 & 10.02 &10.03 \\ $g_{\omega}$ &
12.62 & 12.64 & 12.62 & 12.63 &12.62 \\\hline $b_2$ &-10.53 & -7.10 &
-7.00 & -6.80& -6.83 \\ $b_3$ &-28.00 & -3.80 & 0.00 &3.80& 20.00
\\\hline $c_3$ & 0.29 & 56.77 & 70.77 & 84.71& 159.70 \\\hline\hline
\end{tabular}\\
\end{table}

\begin{figure*}
\epsfig{figure=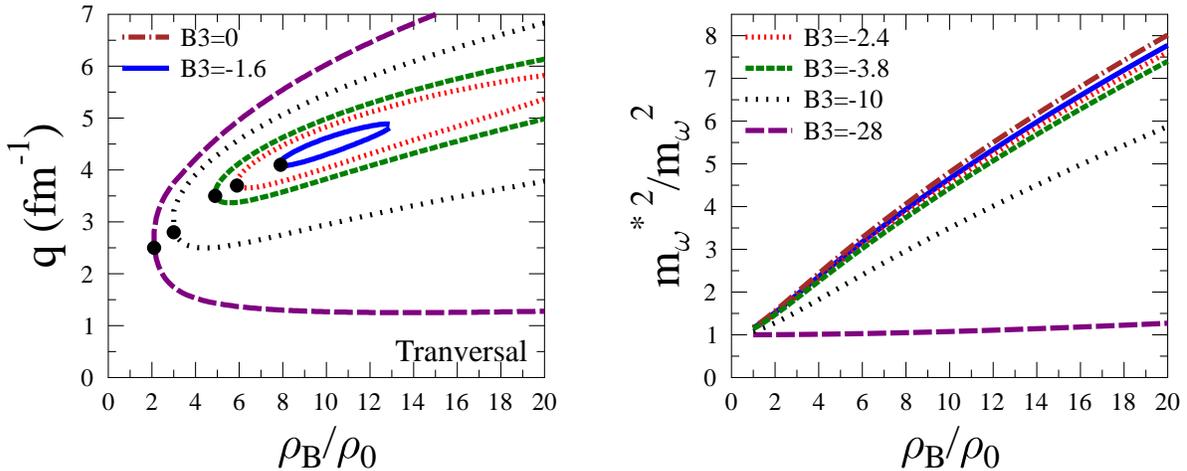, width=18cm}
\caption{The onset of transverse instability (in the left panel) 
and  effective omega meson mass (in the right panel) 
for some test parameter sets. The dot markers show their critical densities.}
\label{t_instab_trmf}
\end{figure*}

\begin{figure*}
\epsfig{figure=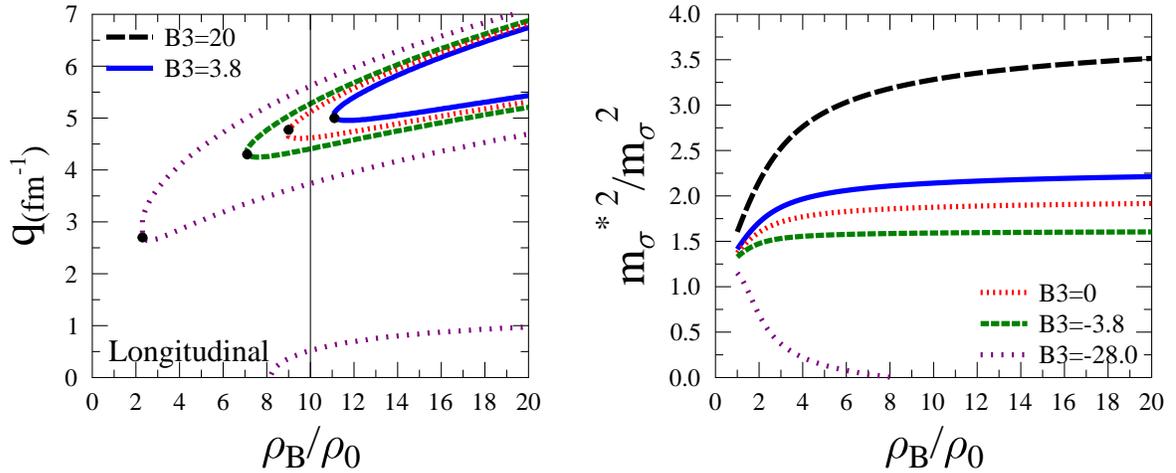, width=18cm}
\caption{The onset of longitudinal instability (in the left panel) 
and  effective sigma meson mass (in the right panel) for some 
test parameter sets. The dot markers show their critical densities.}
\label{l_instab_trmf}
\end{figure*}

\begin{figure}
\epsfig{figure=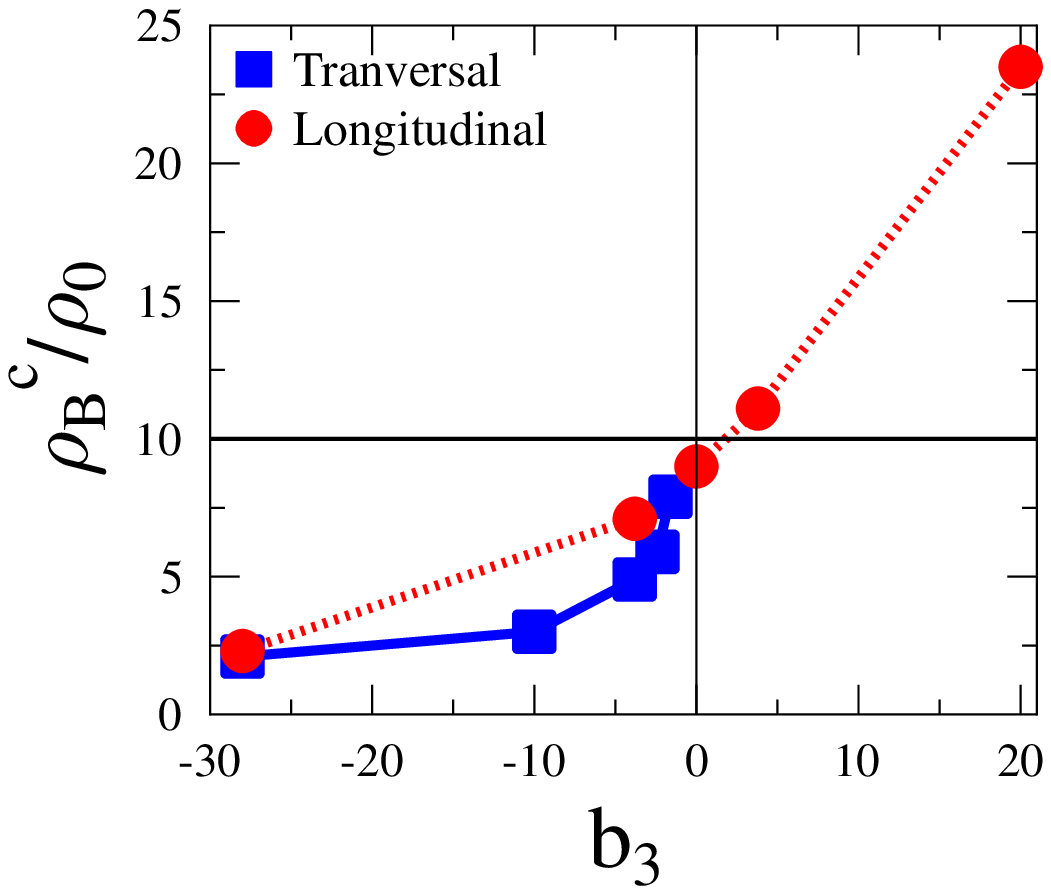, height=8cm}
\caption{The ratio between critical and saturation densities as 
a function of the $b_3$ parameter for both modes.}
\label{dencr}
\end{figure}

In Fig.~\ref{crit_mass_trmf}, the ratio of the critical effective
meson masses to the effective meson masses is shown. The connection of
the $b_3$ parameter set with the appearance of instabilities is
clearly captured.  The explicit relation of the onset of transverse
instability with effective omega meson mass can also be seen in
Fig.~\ref{t_instab_trmf} and the one of the longitudinal mode is shown
in Fig.~\ref{l_instab_trmf}. Fig.~\ref{t_instab_trmf} shows the
correlation between transverse instability and the effective mass of
the omega meson which depends strongly on the vector nonlinear
term. By comparing Fig.~\ref{dencr} with the values of $c_3$ and $b_3$
in Table~\ref{tab:param_trmf}, it is clear that every parameter set
with $c_3>71$ does not show a transverse instability region at high
density. To attain this value for $c_3$, it seems that we need $b_3$
close to or larger than zero.


On the other hand, we can see from Fig.~\ref{l_instab_trmf} that the
longitudinal instability can be related to the effective sigma meson
mass which strongly depends on scalar nonlinear terms. The
longitudinal instability cannot fully disappear, but can be pushed to
unphysical densities by taking $b_3$ relatively large and positive,
for example, $b_3\ge20$, which it corresponds to the value of $c_3\ge
160$. In this case, the onset of instability can be pushed to
$\rho_c\ge23\,\rho_0$ . For high-density applications (neutron stars,
supernova matter, etc.) the maximum density is $10\rho_0$, with
$\rho_0$ the SNM saturation density. This shows that for the purpose
of these applications the parameter set with $b_3>0$ may be
sufficient. For negative but relatively large absolute values of
$b_3$, an additional onset of longitudinal instability is
generated. This appears due to the effective sigma meson mass becoming
imaginary after reaching a certain critical density. This instability
appears if $b_3\le -28$. These results can be also directly seen from
Fig.~\ref{crit_mass_trmf}. Of course for arbitrary standard RMF
parameter sets, its value can be relatively lower or higher, depending
on how large the value of $b_2$ is. Once again, we need to emphasize
that the critical nonlinear parameters are obtained using ``standard''
$g_\omega$ and $g_\sigma$, Therefore for parameter sets with
$g_\omega$ and $g_\sigma$ deviating appreciably from these, their
critical nonlinear parameters can be smaller or larger.

The last finding here is that the parameter sets which are stable at 
high densitiest are  consistent with the experimental data of
Ref.~\cite{Daniel} and the microscopic calculation of Ref.~\cite{Apr}
with respect to their SNM EOS.  

%%%%%%%%%%%%%%%%%%%%%%%%%%%%%%%%%%%%%%%%%%%%%%%%%%%%%%%%%%%%%%%%%%%%%%%%%
\section{Analyzing the stability of some relativistic mean field models}
\label{sec_stability}

Here we will use the criteria defined in the previous section to
analyze more systematically the high density longitudinal and
transverse instabilities of some RMF models. The onset of
instabilities of some parameter sets have been reported in
Ref.~\cite{AMBM07}, but the discussion there was not quite
robust. Therefore, in this work, we need to elucidate one point that
was not treated properly in Ref.~\cite{AMBM07}, namely that both
instabilities are determined by the interplay between critical and
actual effective meson masses (Eq.~(\ref{eq:requr})). To comprehend
our previous results~\cite{AMBM07}, now we also enlarge the number of
parameter sets used.

\subsection{Standard relativistic mean field models}
\begin{figure*}
\epsfig{figure=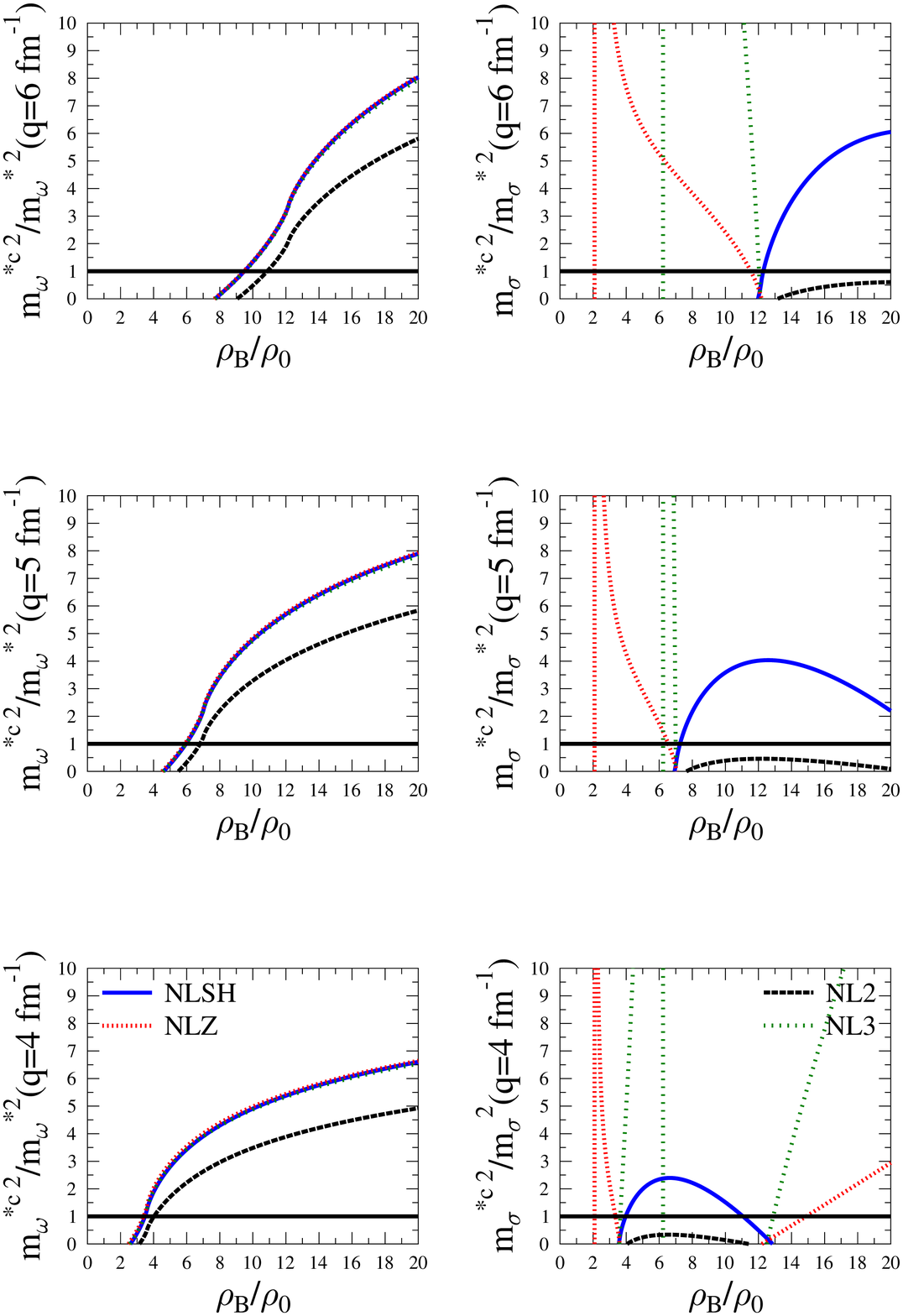, width=16cm}
\caption{ $m_\sigma^{*~2~c}/m_\sigma^{*2}$ and
$m_\omega^{*~2~c}/m_\omega^{*2}$ for some S-RMF parametrizations 
in Table~\ref{tab:params_srmf} as functions of $\rho_B/\rho_0$, with
$q$ varying as indicated.}
\label{crit_mass_srmf}
\end{figure*}

First, we start with S-RMF. We select four parameter sets of this
model as representative, i.e., NLZ~\cite{Rufa}, NL3~\cite{Lala},
NLSH~\cite{Sharma} and NL2~\cite{PGR1}. The isoscalar parameters of
these parameter sets can be seen in Table~\ref{tab:params_srmf}. It
can be seen that $g_\sigma$ and $g_\omega$ are of similar magnitude
for NLZ, NL3, and NLSH, but the ones of NL2 are significantly
smaller. Therefore, the parameter sets NLZ, NL3, and NLSH have similar
$\sigma$ and $\omega$ critical effective masses while those of NL2 are
smaller. Because NL2 has $\sigma$ and $\omega$ critical effective
masses which are smaller than for the others and a positive value of
the $b_3$ parameter, the longitudinal instability of this parameter
set disappears in the range of densities and $q$ values used. On the
other hand NLSH, having a quite large $b_2$ parameter and a moderate
$b_3$ parameter, its sigma meson effective mass is larger compared to
the ones of NLZ and NL3. Therefore, its instability regions are
narrower and have no fluctuations. All parameter sets of S-RMF have a
transverse instability. At relatively small $q$ ($\le$ 4 $\rm
fm^{-1}$), they start to appear in quite low densities. These facts
are clearly depicted in Fig.~\ref{crit_mass_srmf}.

\begin{figure*}
\epsfig{figure=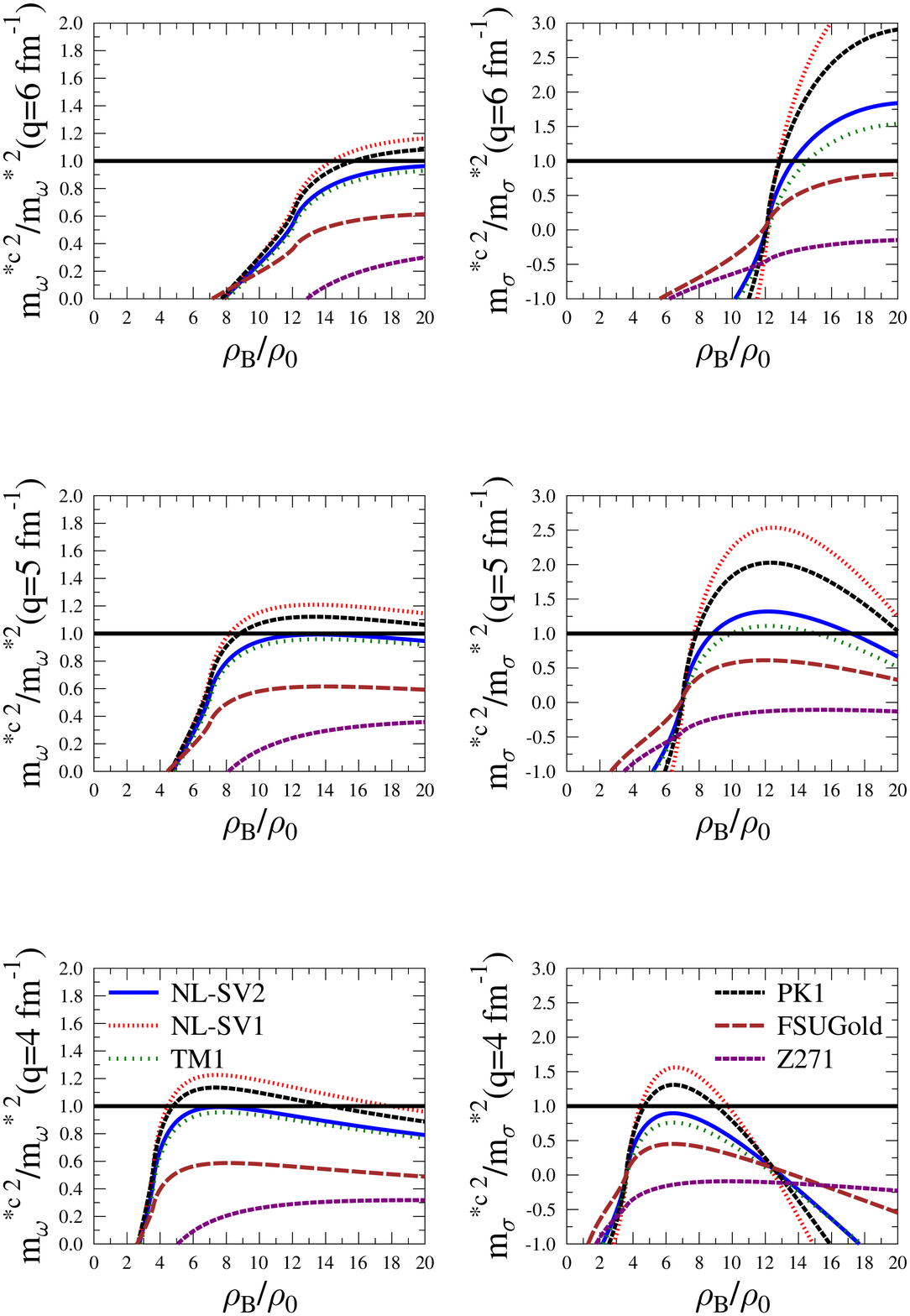, width=16cm}
\caption{ $m_\sigma^{*~2~c}/m_\sigma^{*2}$ and
$m_\omega^{*~2~c}/m_\omega^{*2}$ for  some V-RMF parametrizations 
in Table~\ref{tab:params_vrmf} as functions of $\rho_B/\rho_0$ and
with $q$ varying as indicated.}
\label{crit_mass_vrmf}
\end{figure*}

Next, we will discuss the V-RMF model. Here, we investigate the
parameter sets NLSV1 and NLSV2 of Ref.~\cite{Sharma2},
TM1~\cite{Toki}, PK1~\cite{Long}, Z271~\cite{Pieka1}, and
FSUGold~\cite{Pieka2}. The isoscalar parameters of these parameter
sets can be seen in Table~\ref{tab:params_vrmf}. Due to the small
values of the $g_\sigma$ and $g_\omega$ coupling constants for the
Z271 parameter set, it has relatively small $\sigma$ and $\omega$
critical effective masses, so that even though its $c_3<71$, the
parameter set Z271 does not have a transverse instability region. On
the other hand, for the TM1, NLSV2, and FSUGold parameter sets, their
tranversal instabilities disappear due to the fact that these
parameter sets have $c_3>71$.  In the V-RMF model, it is very clear
reflected that, if $b_3$ becomes smaller then its longitudinal
instability region becomes larger. This fact can be understood by
comparing Fig.~\ref {crit_mass_vrmf} with the $b_3$ values in
Table~\ref{tab:params_vrmf}.
\begin{table}
\centering
\caption {Isoscalar parts of some S-RMF model parameter sets. }
\begin{tabular}{c c c c c}
\hline\hline ~Parameter~~&~NLZ~  & ~NL3~   &~NLSH~ & ~NL2~ \\\hline
$g_{\sigma}$       & 10.05 & 10.22   & 10.44 & 9.11 \\
$g_{\omega}$       & 12.91 & 12.86   & 12.95 & 11.49 \\\hline
$b_2$              &-13.51 & -36.12  & -6.91 & -2.30\\
$b_3$              &-40.22 & -10.43  & -15.83& 13.78\\
$c_3$              & 0     & 0       &0      & 0\\\hline\hline
\end{tabular}\\
\label{tab:params_srmf}
\end{table}
\begin{table}
\centering
\caption {Isoscalar parts of some V-RMF model parameter sets. }
\begin{tabular}{c c c c c c c}
\hline\hline ~Parameter~~&~NLSV1  & PK1   &TM1 & NLSV2 &Z271 &FSUGold \\\hline
$g_{\sigma}$       & 10.12 & 10.32  & 10.03 & 10.32 &7.03  &10.59\\
$g_{\omega}$       & 12.73 & 13.01  & 12.61 & 12.88 &8.41  & 14.30\\\hline
$b_2$              &-9.24  & -8.17  & -7.23 & -6.86& -5.43  & -4.28\\
$b_3$              &-15.39 & -10.00 & 0.62  &0.37&  63.69 &49.93\\\hline
$c_3$              & 41.01 & 55.65  & 71.33 & 72.39& 49.94  &418.39\\\hline\hline
\end{tabular}\\
\label{tab:params_vrmf}
\end{table}

In general, the problem of the transverse part of S-RMF maybe can
remedied by using very large bare omega meson masses but this still
leaves the problem in the longitudinal part that should be fixed. Many
parameter sets of the S-RMF model have negative but relatively large
absolute value of $b_3$. This fact leads to the situation that after
reaching a certain density their effective omega meson mass becomes
imaginary. This effect, as mentioned previously, appears as wild
fluctuations in the critical effective sigma meson
mass~\cite{AMBM07}. The presence of the $c_3$ parameter in the V-RMF
model leads to the point that some parameter sets of this model are
free of transverse instability while for some the critical density
for longitudinal instability can be pushed to an unphysical
region. Thus the presence of $c_3$ larger than a certain value can
indeed remedy both instabilities.

\subsection{Relativistic mean field model with mixing nonlinear terms}
\begin{figure*}
\epsfig{figure=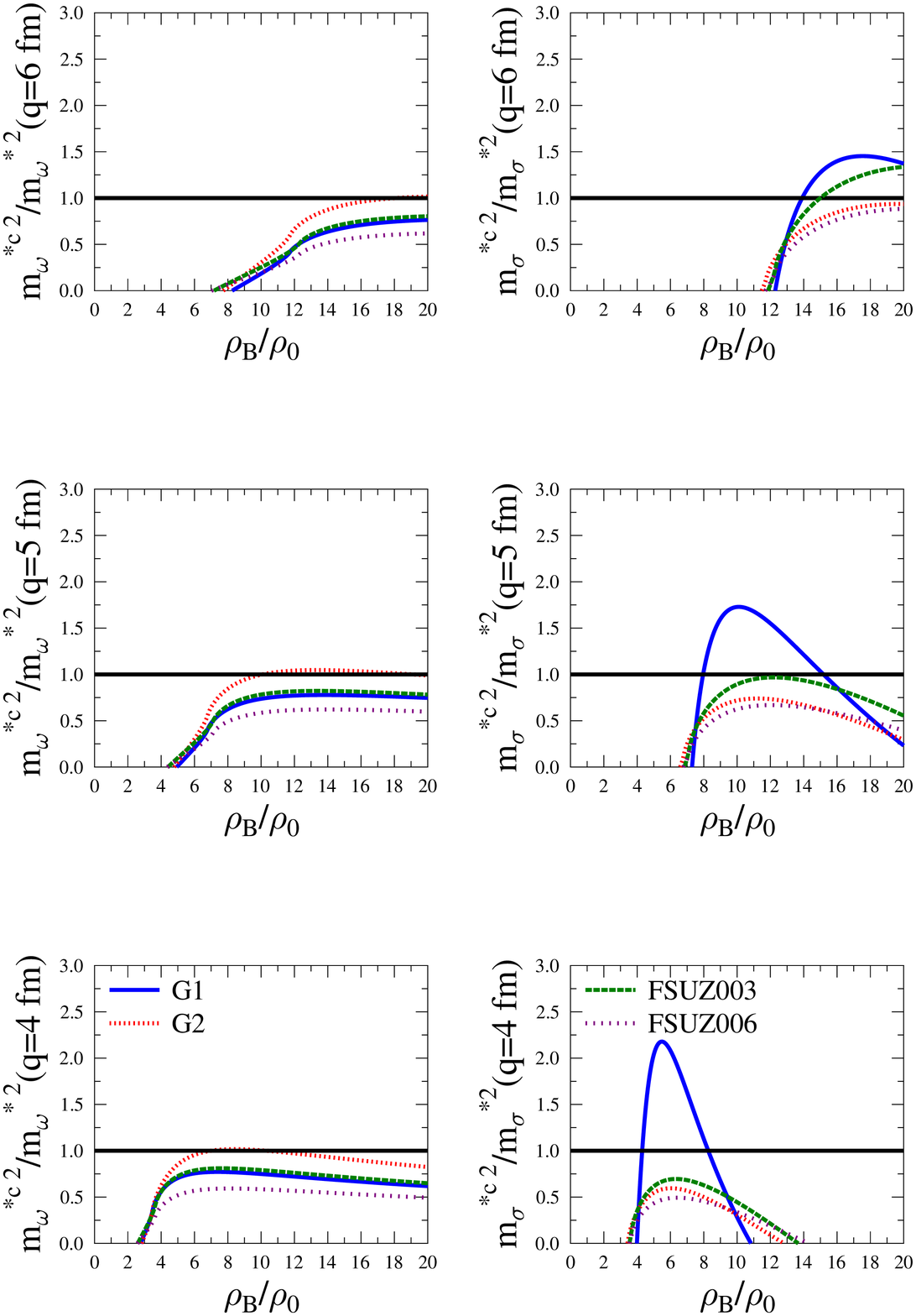, width=16cm}
\caption{ $m_\sigma^{*~2~c}/m_\sigma^{*2}$ and
$m_\omega^{*~2~c}/m_\omega^{*2}$ 
of some E-RMF parameterizations of Table~\ref{tab:params_ermf} as
functions of $\rho_B/\rho_0$ and for some $q$ values as indicated.}
\label{crit_mass_ermf}
\end{figure*}

In this subsection, we discuss in more detail than in our previous
work~\cite{AMBM07}, the effect of the mixing nonlinear terms of some
parameter sets of RMF model (E-RMF) that exist in the
literature. The presence of these terms~\cite{AMBM07} provides
additional terms in the energy density $\varepsilon$
(Eq.~(\ref{eq:sd_edens})) with $- d_2 \sigma V_0^2 - \frac{1}{2} d_3
\sigma^2 V_0^2$, and in the longitudinal dielectric function $\epsilon_L$
(Eq.~(\ref{eq:die})) should be added with $4 d_{SV}^L \Pi_{SV}$, where
$\Pi_{SV} \equiv \Pi_{M}+2 d_{SV}^L\Pi_M^2+ 2d_{SV}^L\Pi_S
\Pi_{00}$. The meson propagators therefore take the following forms
\bea
d_S&=&\frac{g_{\sigma}^2}
{q^2+m^{*~2}_{\sigma}+\Delta_{\sigma \omega}^2{(q^2+m^{*~2}_{\omega})}^{-1}},\nonumber\\
d_V^L&=&\frac{g_{\omega}^2}{q^2+m^{*~2}_{\omega}+\Delta_{\sigma \omega}^2{(q^2+m^{*~2}_{\sigma})}^{-1}},\nonumber\\
\eea 
and now the scalar-vector coupling propagator is nonzero, i.e.,
\bea
d_{SV}^L=\frac{g_{\omega}g_{\sigma} \Delta_{\sigma
\omega}}{(q^2+m^{*~2}_{\omega})(q^2+m^{*~2}_{\sigma})+\Delta_{\sigma
\omega}^2},
\eea
where the effective meson masses become 
\bea
m_\sigma^{*~2}&=& \frac{\partial^2 \epsilon}{\partial^2 \sigma}=m_\sigma^2+2 b_2 \sigma+3 b_3 \sigma^2- d_3 V_0^2 ,\nonumber\\ m_\omega^{*~2}&=& -\frac{\partial^2 \epsilon}{\partial^2 V_0}=m_\omega^2 + 3 c_3  V_0^2+2 d_2 \sigma+d_3 \sigma^2,\nonumber\\ \Delta_{\sigma \omega}&=&
-(\partial^2\epsilon/\partial \sigma \partial V_0)=2 d_2 V_0+2d_3 \sigma V_0.
\label{eq:meseffmassmix}
\eea
These terms also modify $m_\sigma^{*~2~c}$ in Eq.~(\ref{eq:msqmass}) into
\begin{widetext}
\bea
m_\sigma^{*~2~c}&=&-\[\frac{2 g_{\sigma}^2 (\Pi_S \Pi_T-\Pi_M^2+\Pi_S\Pi_{00})(1+\lambda_1)}{( \Pi_T+\Pi_{00})(1+\lambda_2)}+q^2\],
\label{eq:msqmassmix}
\eea
\end{widetext}
while the corrections due to the mixing nonlinear terms are represented by 
\begin{widetext}
\bea
\lambda_1 &=& \frac{1}{(\Pi_S \Pi_T-\Pi_M^2+\Pi_S\Pi_{00})}\(\delta_{\sigma \omega}[\Pi_M^2-\Pi_S\Pi_{00}]\),\nonumber\\
\lambda_2 &=& \frac{1}{(\Pi_T+\Pi_{00})}\(\delta_{\sigma \omega}\Pi_T + 
\frac{ 4 g_{\omega}g_{\sigma}\Pi_T  \delta_{\sigma \omega}}{\Delta_{\sigma \omega}}\[\Pi_M+ \frac{ 2 g_{\omega}g_{\sigma} \delta_{\sigma \omega}}{\Delta_{\sigma \omega}}(1-\delta_{\sigma \omega})[\Pi_M^2-\Pi_S\Pi_{00}]\]\) ,\nonumber\\
\eea
\end{widetext}
with
\bea
\delta_{\sigma \omega}=\frac{\Delta_{\sigma
\omega}^2}{(q^2+m^{*~2}_{\omega})(q^2+m^{*~2}_{\sigma})}.
\eea
It is clear that if $\lambda_1=\lambda_2=0$, these terms do not affect
the critical sigma meson mass. If both $\lambda$'s are not equal to
zero , the critical sigma meson mass reduction or increase depends on
the sign and values of both parameters. On the other hand, the actual
effective masses of sigma and omega mesons still depend explicitly on
the nonlinear mixing terms.
\begin{figure*}
\epsfig{figure=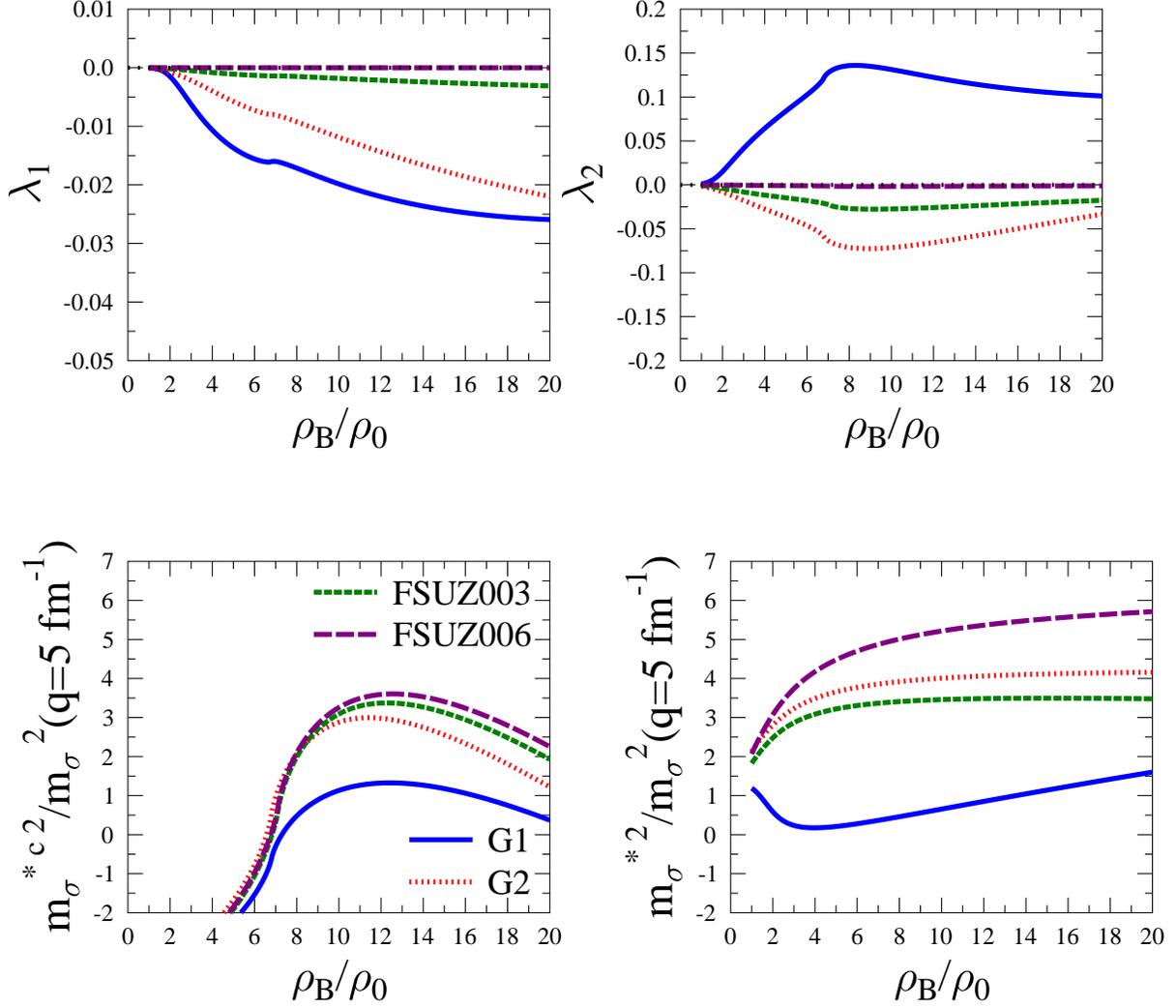, width=18cm}
\caption{ $\lambda_1$ (in left upper panel), $\lambda_2$ (in the right upper panel),
$m_\sigma^{*~2~c}/m_\sigma^{2}$ (in the left lower panel) and
$m_\omega^{*~2}/m_\omega^{2}$ (in the right lower panel) for some
E-RMF parameterizations and $q= 5 {\rm fm}^{-1}$.}
\label{lambda}
\end{figure*}
As the representatives of the E-RMF model, the parameter sets G1 and
G2 from Ref.~\cite{Furnstahl96}, as well as FSUZ003 and FSUZ006 from
Ref.~\cite{Kumar} are used. The isoscalar parameters of these
parameter sets can be seen in Table~\ref{tab:params_ermf}.  The
$\lambda_1$, $\lambda_2$, $m_\sigma^{*~2~c}/m_\sigma^{2}$ and
$m_\omega^{*~2}/m_\omega^{2}$ of the representative parameter sets for
the case $q= 5{\rm fm}^{-1}$ are shown in Fig.~\ref{lambda}. In this
case, the maximal values are $\lambda_1\approx2.5$\% and
$\lambda_2\approx 15$\% of the total contributions. Both are obtained
using the G1 parameter set. This implies that the mixing nonlinear
terms do not significantly affect the critical effective mass of the
sigma meson. This can be seen from the lower left panel of
Fig.~\ref{lambda}. $m_{\sigma}^{*~2~c}/m_{\sigma}^2$ of the G1
parameter set is significantly lower than those of the other parameter
sets, merely because it produces significantly smaller $g_{\sigma}$
and $g_{\omega}$ compared to them (see Table
~\ref{tab:params_ermf}). However, the significant effect of mixing
nonlinear terms in the actual effective mass of the sigma meson can be
seen in the right panel of Fig.~\ref{lambda}.
\begin{table}
\caption {Isoscalar parts of some E-RMF model parameter sets.}
\begin{tabular}{c c c c c}
\hline\hline ~Parameter~~&~G1~  & ~G2~   &~FSUZ003~ & ~FSUZ006~\\\hline
$g_{\sigma}$       & 9.87 & 10.49  & 10.76 & 11.02 \\
$g_{\omega}$       & 12.13 & 12.76 & 14.11 & 14.66 \\\hline
$b_2$              &-15.09 &-24.89 & -9.72 & -4.65\\
$b_3$              &-47.69 & 3.56  & 21.80 & 60.28\\\hline
$c_3$              & 86.41 & 71.71 & 198.25& 462.64\\\hline
$d_2$              & -1.15 &-11.26 &-2.21  & -0.17\\
$d_3$              & -32.56& 4.19  & 11.69 & 1.38\\\hline\hline
\end{tabular}
\label{tab:params_ermf}
\end{table}
For the transverse mode, it is clear that G2 has an instability
region because its $c_3$ is critical, i.e., produces the maximal
effective mass of the omega meson, equal to the critical value. If the
parameter $c_3$ is larger than this value like for the other E-RMF
parameter sets, the instability disappears. For the longitudinal mode,
the appearance of instability is merely determined by the effective
sigma meson mass. The combination of $b_2$, $b_3$, and $d_3$ of each
parameter set leads to $m_{\sigma ~{\rm G1}}^{*2}<m_{\sigma ~{\rm
FSUZ003}}^{*2}<m_{\sigma ~{\rm G2}}^{*2}<m_{\sigma ~{\rm
FSUZ006}}^{*2}$ (right lower panel of Fig.~\ref{lambda}). Thus the
critical density $\rho_c^{\rm G1}<\rho_c^{\rm FSUZ003}<\rho_c^{\rm
G1}<\rho_c^{\rm FSUZ006}$. The large and positive value of $b_3$ is
still the way to push the longitudinal instability to unphysical
regions, but in the E-RMF model, the value of the critical $b_3$ need
not be too large, because its contribution can be partly substituted
not only by that of $b_2$ but also by that of $d_3$. These facts can
be understood by comparing Fig.~\ref{crit_mass_ermf} with nonlinear
parameters in Table~\ref{tab:params_ermf}. Another interesting finding
is that if we observe the parameters of G2 in
Table~\ref{tab:params_ermf}, even its $b_3$ is negative and its
absolute value is relatively large, but other contributions from $b_2$
and $d_3$ produce sufficient balancing contributions, avoiding this
parameter set to have an imaginary effective sigma meson mass.

%%%%%%%%%%%%%%%%%%%%%%%%%%%%%%%%%%%%%%%%%%%%%%%%%%%%%%%%%%%%%%%%%%%%%%%%%
\section{Conclusion}
\label{sec_conclu}
Based on the fact that the critical effective masses of the sigma and
omega mesons ($m_\sigma^{*~2~c}$ and $m_\omega^{*~2~c}$) depend on the
$g_{\sigma}$ and $g_{\omega}$ coupling constants but are insensitive
to the values of the $b_2$, $b_3$, and $c_3$ parameters, while the
actual effective masses of the sigma and omega mesons
($m_\sigma^{*~2}$ and $m_\omega^{*~2}$) have opposite properties, we
derive criteria to estimate the values of isoscalar nonlinear terms of
the standard RMF model that produce a stable EOS at high
densities. The minimal requirement on the standard RMF model free from
transverse mode instability at high densities is that the quartic
vector nonlinear parameter should exist and have a value of $c_3>71$. 
This value corresponds to the quartic scalar nonlinear parameter
$b_3$ close to or larger than zero. The longitudinal mode instability
cannot fully disappear but can be pushed to unphysical densities by
taking $b_3$ relatively large and positive, for example, $b_3\ge20$,
where in our chosen test parameter set, it corresponds to
$c_3\ge160$. 
In this case, the critical density for the instability region can
be pushed to $\rho_c\ge23~\rho_0$ . In usual applications, for
examples neutron stars, supernova matter, etc., the density remains
below $10\rho_0$, so that for these applications it seems sufficient
to choose a parameter set with $b_3$ slightly lower than 20 is
sufficient. We have also found that the parameter set of the standard
RMF model with a stable nuclear matter EOS at high density is
consistent with the experimental data of Ref.~\cite{Daniel} and the
microscopic calculation of Ref.~\cite{Apr}.

The criteria has been used to systematically analyze  the high density
longitudinal and transverse modes instabilities of standard RMF
models.  We have shown that both parameter sets unstable at high
densities and stable ones exist. The reason behind this
fact has been explained quantitatively and  the crucial role of the
isoscalar vector nonlinear term has also been clearly
demonstrated. The effect of the mixing nonlinear terms of some
parameter sets of the RMF model (E-RMF) has been also studied. Due to the
contribution of these parameters to the actual effective
masses of the sigma and omega mesons ($m_\sigma^{*~2}$ and
$m_\omega^{*~2}$) are changed, while the
critical effective masses of sigma  meson ( $m_\sigma^{*~2~c}$) are
affected only weakly. The
stability of both modes  can be achieved more easily compared to
V-RMF because the thresholds of the critical values of  $c_3$ and $b_3$
parameters of this model can be lower than those of  V-RMF.   
%%%%%%%%%%%%%%%%%%%%%%%%%%%%%%%%%%%%%%%%%%%%%%%%%%%%%%%%%%%%%%%%%%%%%%%%%

%%%%%%%%%%%%%%%%%%%%%%%%%%%%%%%%%%%%%%%%%%%%%%%%%%%%%%%%%%%%%%%%%%%%%%%%%
\section*{ACKNOWLEDGMENT}
A. S. would like to thank M. M. Sharma for useful exchange of ideas
connected with this work.  A.S. also acknowledges financial support from DAAD and Universitas Indonesia. He is grateful for the kind hospitality during his stay in Institut f\"ur Theoretische Physics at Universit\"at Frankfurt.
%%%%%%%%%%%%%%%%%%%%%%%%%%%%%%%%%%%%%%%%%%%%%%%%%%%%%%%%%%%%%%%%%%%%%%%%%


\begin {thebibliography}{50}
\bibitem{Lala09} G. A. Lalazissis, S. Karatzikos, R. Fossion, D. Pena Artega, and P. Ring, 
\Journal{\PLB}{671}{36}{2009}.
\bibitem{Sharma08} M. M. Sharma, 
\Journal{\PLB}{666}{140}{2008}.
\bibitem{Kumar} R. Kumar, B. K. Agrawal and S. K. Dhiman,
\Journal{\PRC}{74}{034323}{2006}.
\bibitem{Pieka2} B. G. Todd-Rutel and J. Piekarewicz,
\Journal{\PRL}{95}{122501}{2005}.
\bibitem{Klahn} T. Klaehn $\it{et.~al}$,
\Journal{\PRC}{74}{035802}{2006}.
\bibitem{Sagert} I. Sagert, M. Wietoska, J. Schaffner-Bielich and C. Sturm,
\Journal{\JPG}{35}{014053}{2008}.
\bibitem{Pieka07} J. Piekarewicz,
\Journal{\PRC}{76}{064310}{2007}.
\bibitem{AMBM07} A. Sulaksono, T. Mart, T. J. Buervenich and J. A. Maruhn,
\Journal{\PRC}{76}{0431301(R)}{2007}.
\bibitem{Horo3} K. Lim and C. J. Horowitz,
\Journal{\NPA}{501}{729}{1989}.
\bibitem{Fri} B. L. Friman and P. A. Henning, 
\Journal{\PLB}{206}{579}{1988}.
\bibitem{PGR2} H.-G. Doebereiner and P.-G. Reinhard, 
\Journal{\PLB}{227}{305}{1989}.
\bibitem{Daniel}P. Danielewicz, R. Lacey and W.G. Lynch, 
\Journal{Science}{298}{1592}{2002}.
\bibitem{Apr}A. Akmal, V.R. Pandharipande and D.G. Ravenhall, 
\Journal{\PRC}{58}{1804}{1998}.
\bibitem{Rufa} M. Rufa, P.-G. Reinhard, J. A. Maruhn, W. Greiner and M. R. Strayer, 
\Journal{\PRC}{38}{390}{1988}.
\bibitem{Lala}G. Lalazissis, J. K\"onig and P. Ring,
\Journal{\PRC}{55}{540}{1997}.
\bibitem{Sharma} M. M. Sharma, M. A. Nagarajan and P. Ring,
\Journal{\PLB}{312}{377}{1993}.
\bibitem{PGR1} P.-G. Reinhard, 
\Journal{Rep. Prog. Phys}{52}{439}{1989}.
\bibitem{Sharma2} M. M. Sharma, A. R. Farhan and S. Mythili,
\Journal{\PRC}{61}{054306}{2000}.
\bibitem{Toki}Y. Sugahara and H. Toki,
\Journal{\NPA}{579}{557}{1994}.
\bibitem{Long} W. Long, J. Meng, N. VanGiai  and S.-G. Zhou,
\Journal{\PRC}{69}{034319}{2004}.
\bibitem{Pieka1} C. J. Horowitz and J. Piekarewicz,
\Journal{\PRL}{86}{5647}{2001}.
\bibitem{Furnstahl96} R. J. Furnstahl, B. D. Serot and H. B. Tang,
\Journal{\NPA}{598}{539}{1996}; \Journal{\NPA}{615}{441}{1997}.
\end{thebibliography}
\end{document}